\documentclass[prd, onecolumn,showpacs,nofootinbib,preprintnumbers,superscriptaddress]{revtex4-2}
\usepackage{amsmath}\usepackage{amsfonts}\usepackage{graphicx}\usepackage{hyperref}\usepackage{mathbbol}
\def\be{\begin{equation}}  \def\ee{\end{equation}}   \def\bea{\begin{eqnarray}}    \def\eea{\end{eqnarray}}  \def\no{\nonumber}
\def\e{{\epsilon}}  \def\ve{{\varepsilon}}   \def\d{{\rm d}}   \def\m{{\rm m}}      \def\k{\kappa}          \def\p{{\rm P}}
\def\f{\frac}    \def\l{\left}   \def\r{\right}

\begin{document}
\title{Cosmology in holographic non-minimal derivative coupling theory: constraints from inflation and variation of gravitational constant}
\date{\today}

 \author{Phichayoot Baisri}\email{phichayootb62@nu.ac.th}
 \affiliation{The Institute for Fundamental Study ``The Tah Poe Academia Institute", Naresuan University, Phitsanulok 65000, Thailand}
  \affiliation{NAS, Centre for Theoretical Physics \& Natural Philosophy, Mahidol University, Nakhonsawan Campus,  Phayuha Khiri, Nakhonsawan 60130, Thailand}
  \author{Burin Gumjudpai}\email{Corresponding: burin.gum@mahidol.ac.th}
  \affiliation{NAS, Centre for Theoretical Physics \& Natural Philosophy, Mahidol University, Nakhonsawan Campus,  Phayuha Khiri, Nakhonsawan 60130, Thailand}
\author{Chonticha Kritpetch}\email{chonticha.kr@up.ac.th}
  \affiliation{NAS, Centre for Theoretical Physics \& Natural Philosophy, Mahidol University, Nakhonsawan Campus,  Phayuha Khiri, Nakhonsawan 60130, Thailand}
   \affiliation{High Energy Physics and Cosmology Research Group, School of Science, University of Phayao, Phayao 56000, Thailand}
 \author{Pichet Vanichchapongjaroen}\email{pichetv@nu.ac.th}
  \affiliation{The Institute for Fundamental Study ``The Tah Poe Academia Institute", Naresuan University, Phitsanulok 65000, Thailand}

\begin{abstract}
We consider a cosmological model of non-minimal derivative coupling (NMDC) to gravity with holographic effect from Bekenstein-Hawking entropy
using Hubble horizon IR cutoff.  Holographic parameter $c$ is considered constant with value in a range, $0 \leq c < 1$.  The NMDC effect is  considered either as modification in kinetic scalar field or in gravitational constant.
The NMDC effect allows gravitational constant to be time-varying. Since the NMDC effect is cosmological, definition of holographic density should include time-varying part of the gravitational constant.
The NMDC part reduces strength of gravitational constant for $\k > 0$  and opposite for $\k < 0$. The holographic part enhances gravitational strength. Slow-roll parameters are derived.
We use spectral index and tensor-to-scalar ratio to test the model against CMB constraint.
Number of e-folding is chosen to be $N \geq  60$.
Power-law scalar potentials, $V = V_0 \phi^n $ with $n = 2, 4$, and exponential potential, $V = V_0 \exp{(-\beta \phi)}$ are considered.  Combined parametric plots of $\k$ and $\phi$ show that the allowed regions of the power spectrum index and of the tensor-to-scalar ratio are not overlapping. NMDC inflation is ruled out and the holographic NMDC inflation is also ruled out for $0 < c < 1$.
NMDC significantly changes major anatomy of the dynamics, i.e. it gives new late-time attractor trajectories in acceleration regions.   The holographic part clearly affects pattern of trajectories. However, for the holographic part to affect shape of the acceleration region, the NMDC field must be in presence.   To constrain the model at late time,
variation of gravitational constant is considered. Gravitational-wave standard sirens and supernovae data give a constraint, $\dot{G}/G|_{t_0} \lesssim 3\times10^{-12} \,  \text{year}^{-1}$ \cite{Zhao:2018gwk} which, for this model, results in
$
10^{-12} \,  \text{year}^{-1}   \,  \gtrsim  \,      {- \kappa} \dot{\phi}\ddot{\phi}/{M^2_{\p}}\,.
$
Positive $\k$ is favored and greater $c^2$  results in lifting up lower bound of $\k$.
\end{abstract}

\date{\today}
\vskip 1pc
\maketitle \vskip 1pc
%%%%%%%%%%%%%
%%%%%%%%%%%%%%%%%%%%%%
\section{Introduction}  \label{sec_intro}

Over the last decades, numerous approaches  (as reviewed in \cite{CopelandDynamics of dark energy, Paddy2006, DEBook, VF,CL,CF, odin,en, Ishak:2018his} and many references therein) have been purposed to clarify cosmological puzzles, e.g. present acceleration with $w \approx -1$   \cite{Amanullah2010, Astier:2005qq, Goldhaber:2001a, Perlmutter:1997zf, Perlmutter:1999a, Riess:1998cb, Scranton:2003, Tegmark:2004} and graceful exit of the inflation confirmed by CMB \cite{WMAP9, Aghanim:2019ame, Planck:2018vyg}.   One approach to explain  late-time acceleration is dark energy, a hypothetical source of repulsive pressure with equation of state $w < -1/3$.  These dark energy models  include cosmological constant and dynamical  scalar fields. Inclusion of dark energy is to add extra degrees of freedom to the matter Lagrangian.  The other major idea is to make some modifications to gravitational sector. The geometry sector is modified in form of function of Ricci scalar, Ricci tensor and Riemann tensor in this approach \cite{Carroll2004}.  Many of other hypotheses are of mixed types such that couplings among barotropic fluid, scalar sector and geometry sector are allowed.  Scalar-tensor theories have rich implications of  these couplings \cite{BD1961, Maeda, VF}.   As there is coupling between matter and scalar field, chameleon screening mechanism is required to protect these models from mediating long-range fifth force \cite{charm}.  The coupling function in form of $f(\phi, \phi_{,\mu}, \phi_{,\mu\nu}, \ldots)$ to gravitational sector  are inspired in many theories such as  scalar quantum electrodynamics or  in gravitational theories of which Newton's constant is a function of the density \cite{Amendola1993}. It was shown that
other non-minimal derivative coupling (NMDC) to gravity terms apart from
 $R \phi_{,\mu}\phi^{,\mu}$
and  $R^{\mu\nu} \phi_{,\mu} \phi_{,\nu}$  are unnecessary \cite{Capozziello:1999xt}.  These two terms exist  in  lower energy limits of higher dimensional
theories or Weyl anomaly of $\mathcal{N} = 4$ conformal supergravity
\cite{Liu:1998bu, Nojiri:1998dh}.  It is interesting that NMDC gravity in form of  $\kappa_1 R \phi_{,\mu}\phi^{,\mu}$  and
 $\kappa_2 R^{\mu\nu} \phi_{,\mu} \phi_{,\nu}$ usually gives de-Sitter expansion \cite{Capozziello:1999uwa}  and they  can be further modified to $\kappa \equiv   \kappa_2  =  -2 \kappa_1$
 \cite{Granda:2010hb, Granda:2010ex, Granda:2011zk} resulting a hint of NMDC term coupling to the Einstein tensor, $G^{\mu\nu}$.
  Both metric tensor and the Einstein tensor are divergence-free, therefore the theory is naturally settled    \cite{Sushkov:2009, Saridakis:2010mf, Gao:2010vr, Germani:2010gm, Sushkov:2012, Skugoreva:2013ooa, Koutsoumbas:2013boa, Darabi:2013caa, Germani:2009, Germani:2010hd, Dalianis:2016wpu, Sadjadi:2012zp, Tsujikawa:2012mk, Ema:2015oaa, Jinno:2013fka, Ema:2016hlw, Yang:2015pga, Sadjadi:2010bz, Gumjudpai:2015vio}. Further versions of the theories are those of generalized type of scalar-tensor theories, for examples, galileons \cite{Nicolis:2008in, Deffayet:2009wt, Deffayet:2009mn} and the Fab-Four \cite{Charmousis:2011bf} with at most second-order derivative with respect to its dynamical variables which are the metric tensor and the scalar field.
 It is found that the NMDC term is a sub-class of the Horndeski Lagrangian \cite{Horn, Deffayet:2011, Kobayashi} and surely a sub-class of the beyond Horndeski theories such as GLPV theories \cite{GLPV}.

 On cosmological aspects, the coupling $\k$, if allowed to vary or change sign, could enhance or reduce contribution of  the free kinetic term  \cite{Granda:2010hb} which in turn affects power spectral index, tensor-to-scalar ratio, evolution of the equation of state and other observational parameters  \cite{Caldwell:2003vq, Saridakis:2010mf, Skugoreva:2013ooa, Quiros:2017gsu}.
Observational data puts tight constraints on the form of scalar potentials \cite{Tsujikawa:2012mk, Yang:2015pga}, giving criterion conditions for NMDC models to be viable either as a driving force of early inflationary phase or of late-time acceleration phase. In Palatini formalism, the NMDC inflation with quartic power-law potential is disfavored by CMB data \cite{Gumjudpai:2016ioy, Saichaemchan:2017psl, Muhammad:2018dwi}.
NMDC inflation succeeds in resulting in quasi-de Sitter expansion with graceful exits if the field is initially fast-rolling. This happens even without scalar potential. On the other hand, at late time
the theory yields $w \rightarrow -1$   \cite{Sushkov:2009, Bruneton:2012zk, Sushkov:2012}. It is interesting that acceleration is possible in this theory for a power-law potential  $V(\phi) = V_0 \phi^n$  with  $n \leq 2$ for sub-Planckian value of  $\k$ and of $V$ and the acceleration ends with scale factor oscillation  \cite{Sushkov:2012, Skugoreva:2013ooa}. However when  $n > 2$, the expansion encounters the Big Rip singularity. Higgs-like potential and exponential potential have been investigated.  Recent work \cite{Avdeev:2021von} considers $n=1.5$, reasonable (i.e. sub-Planckian) initial field value can give sufficient e-folding number for solving the horizon problem with large value of coupling and a very small scalar mass. In their works, the dynamics is of double inflation scenario, i.e. the kinetic-term driving inflation contributes to some e-folding number followed by potential driven e-folding number. However shortcomings of the NMDC inflation turns up when the tensor-to-scalar ratio is predicted too large.
Moreover, for  Higgs-like potential or any power-law potential with $n \leq 2$, there is no graceful exit  \cite{Matsumoto:2017gnx, Granda:2017dlx}. NMDC inflation with power-law potentials is hence disfavored by CMB data in this setting. Most recent report, \cite{Avdeev:2022ilo} has ruled out the NMDC inflation with power-law potential using $\k < 0$ whereas the sign of $\k$  therein is defined oppositely from ours.

 Quantum gravity supports a compelling principle, the holographic principle proposed by 't Hooft in 1993 \cite{tHooft:1993dmi}. String theory description of the principle was found soon after by Susskind \cite{Susskind}. In this description, conformal field theory on the surface enclosing a volumic region is viewed as hologram of corresponding string theory describing physics in the bulk. This is known as AdS/CFT correspondence \cite{Maldacena:1997re}.  According to the principle, surface area enclosing a bulk region is related to entropy of the region. The entropy is known as Bekenstein-Hawking entropy \cite{Be1, Be2, Be3, Hawking, Hawking2}. This concept introduces entropy limit of any bulk region. The bound is due to  limited number of quantum states on the surface of which area can be sub-divided into smallest unit area in Planck scale, implying limited amount of information that can be contained in the region \cite{Bousso:1999dw, Bousso:2002bh}. If information (number of states) of the bulk region exceeds the entropy bound, the bulk region turns to be a blackhole.  Entropy of a blackhole hence scales with its surface area, $S \sim A/4G$  or  with square of length scale of a blackhole, $L_{\rm bh}^2$.  Therefore an enclosed bulk matter region can not have entropy exceeding that of blackhole with the same volume size, otherwise the bulk matter region becomes a blackhole.  In this picture, a blackhole is viewed as a hologram of the information at the surface of event horizon \cite{Bousso:2002ju}.
 %%%%%%%%%%%%%%%%%%%%%%%%%%%%%%%%%%%%%%%%%%%%%%%%%%%%HERE
 %%%%%%%%%%%%%%%%%%%%%%%%%%%%%%%%%%%%%%%%%%%%%%%%%%%%

  Applying the holographic principle, i.e. the entropy bound, to cosmology, gives a bound value for equation of state $w< 1$ and a requirement of infinite size of the universe \cite{Fischler:1998st}.
 As a result, UV energy scale, $\rho_{\Lambda}$  and IR cosmic length scale, $L$ of vacuum energy are related via its entropy, i.e. $\rho_{\Lambda} \propto S L^{-4}$ \cite{Cohen, Thomas, Horava, Hsu}.  The universe is viewed as a hologram of information on the surface of cosmic boundary. Relation between the UV energy scale and the IR length scale hence results that
 \begin{eqnarray}  \label{1}
\rho_{\Lambda} &=&  \frac{3c^2}{8\pi GL^2},
\end{eqnarray}
 suggesting IR cutoff scale $L$ to the vacuum energy (cosmological constant) density  \cite{Li}. This idea is known as holographic dark energy (HDE) model. The parameter $c$  is a constant $(0 < c  \leq 1)$  and $M^2_{\rm P}  \equiv  (8 \pi G)^{-1} $. When $c = 0$, the holographic effect hence vanishes.
 At this point, the scenario gives hope to solve the cosmological fine-tunning problem. However, if the length scale is Hubble horizon, $L \sim H^{-1}$, it leads to incompatible value of dark energy equation of state. That is, instead of $w < -1/3$, it gives a dust-like, $w \approx 0 $ equation of state. This is hold only for the flat case and there is only $ \rho_{\Lambda} $ as a sole density component in the universe \cite{Hsu}.
 One might try particle horizon,  $     R_{p} = a \int^{t}_{0}  a^{-1}  {\d t'} = a \int^{a}_{0}   (1/{Ha'^{2}}) {\d a'}  $  as a cutoff length scale \cite{Fischler:1998st, Bousso:1999xy} but in the case of late universe, it does not result in cosmic acceleration, i.e. it gives $w > -1/3$ \cite{Li}. In order to satisfy the observational bound of the equation of state for an accelerating universe ($w < -1/3 $), a next trial is to use future event horizon,  $     R_{h} = a \int^{\infty}_{t}  a^{-1}  {\d t'} = a \int^{\infty}_{a}   (1/{Ha'^{2}}) {\d a'}  $  as IR cutoff length scale instead of the particle horizon \cite{Li, Pavon}. This results in acceptable vacuum energy density (as dark energy) and it can solve the cosmic coincidence problem assuming inflationary e-folding number  $N> 60$ \cite{Kim:2007kw}.  Solving of coincidence problem is not a surprise because the holographic energy density depends on cosmological horizon size  which in turn depends on the amount of inflation.     With an arbitrary type of horizon cutoffs, the Casimir energy is found to be proportional to the horizon size hence advocating the holographic principle \cite{Li:2009pm}.
 Phantom crossing is allowed by observations however the phantom crossing in holographic universe can violate the second law of thermodynamics
 \cite{Gong:2006sn}. To evade the problem, one can consider interaction between HDE and dark matter (DM). This is dubbed the interacting holographic dark energy (IHDE) model. With the interaction, the effective equation of state is varying such that the phantom crossing and tracker solution are possible in a range of observational parameters \cite{Pavon, Wang:2005jx,  KimLeeMyung2006, Setare2006, HuLing2006, BergerShojaei2006, LiGuoZhang2006, Setare:2007we, Karwan:2008ig}.   Despite of the additional merit that the IHDE with future event horizon cutoff could help avoiding Big-Rip singularity which exists for $c < 1$  \cite{LiLinWang2008, LiLiWangZhang2009}, the model has been either tightly constrained or disfavored by a number of observations  \cite{WuGongWangAlcaniz2008, FengWangGongSu2007, ZhangLiuZhang2008, LiLiWangWangZhang2009, FengZhang2016}.  Up to this stage, using future event horizon as cutoff length scale in the original HDE model seems to be favored. However, comparing with old high-$z$ objects, age of the universe predicted is younger than those of the old high-$z$ objects unless forcing $h \lesssim 0.56$  \cite{Wei:2007ig}.   IHDE with non-flat geometry is considered to  soften the age problem, even tough tightly constraint.
It is also true that HDE-DM interaction strength plays a key role in alleviating the cosmic age problem. Nevertheless, the model needs to involve with too many free parameters  \cite{Cui:2010dr}.

The original HDE model with future event horizon cutoff is also plagued with causality problem which is the global property of the event horizon itself  \cite{Cai:2007us}. A few year later, it  was argued that such causality violation (using the event horizon cutoff) does not exist \cite{Kim:2012ik}) and soon after, another new HDE model derived from action principle, is proposed. The new model is free from the causality problem  \cite{Li:2012xf}.
To solve the age problem of the original HDE model (without DE-DM interaction), first version of agegraphic holograhphic dark energy (AHDE) model is proposed considering cosmic time as length scale cutoff \cite{Cai:2007us}. Since the first version of the model does not allow  dark energy to evolve from sub-dominant component at early time to dominant component at late-time, the second AHDE version was proposed to cure the problem using conformal time, $ \eta = \int {\d t}/{a} = \int  ({a^{2}H})^{-1} {\d a}$ as cutoff scale \cite{Wei:2007ty, Wei:2007xu}.  In the same year, a proposal of another IR cutoff scale motivated by spacetime curvature, the Ricci scalar \cite{Gao:2007ep},
 $ L = R  \equiv  -6 \left(\dot{H} + 2H^{2}+ {k}/{a^{2}} \right)$  was made. It is known as Ricci holographic dark energy (RHDE).
Early observational constraints as of RHDE model are reported in  \cite{Zhang:2009un}.  Not long after, the AHDE and RHDE models are finally excluded by observations  \cite{LiLiWang2009, XuZhang2016}.  The other interesting IR cutoff is proposed by Granda and Oliveros (G-O) with cutoff length scale, $
L =  \alpha H^{2} + \beta \dot{H}$   where $\alpha$ and $\beta$ are model parameters. The G-O cutoff is clearly a generalized case of the RHDE model, i.e. the $k = 0$ case of the G-O HDE recovers RHDE model. The G-O model is free from causality problem because it depends on $H$ which is local quantity and it can also solve the coincidence problem \cite{Granda:2008dk, Granda:2008tm}. Moreover the universe with G-O HDE model is possible to accommodate old high-$z$ objects \cite{Granda:2009dia}.
It is noted that there are also other proposals of the cutoff scales such as using higher-order derivative of the Hubble parameter as the cutoff scale and etc.
\cite{Chen:2009zzv, Chattopadhyay:2020mqj, Chattopadhyay:2020xx, Chattopadhyay:2020x2}. A model of generalized length scale cutoff is noticed in \cite{Nojiri:2021iko}.
%%%%
%%%%

In original HDE model, two types of the cutoffs,  Ricci scalar and G-O cutoffs are disfavored by observations of expansion combined with growths of perturbation data while future event horizon cutoff can not be rejected by the data \cite{Akhlaghi:2018knk}. These can be cured when  time-varying $c$ parameter is allowed \cite{Malekjani:2018qcz}.
Moreover, transition from early deceleration to late acceleration is more consistent with the model with Hubble scale cutoff with time-varying $c$ \cite{Malekjani:2018qcz}.   Recently in Ref. \cite{Colgain:2021beg}, the original HDE model with future event horizon cutoff is considered against $H(z)$, CMB, BAO and SN Ia data. Turning point problem of Hubble parameter is claimed to violate the Null Energy Condition (NEC). However this is not surprising since the model with $c < 1$  results in phantom equation of state. Moreover, the age problem of the original HDE model with future event horizon cutoff has not yet been solved.   As the holographic length scale should be a horizon of causality, the carrier of causality, i.e. light, should never reach the horizon. Cosmic bulk volume in holographic scenario should be a volume enclosed by a trapped null surface, mimicking the same concept of blackhole's event horizon. In accelerating universe, apparent horizon exists as a trapped null surface. Light moving towards the apparent horizon will never reach the apparent horizon. Therefore, in an accelerating universe,  it is sensible to consider the apparent horizon as holographic cutoff scale.   It is shown that there is connection of the first law of thermodynamics to the Friedmann equation \cite{Cai:2005ra}. Definition of Cai-Kim temperature, $T = 1/(2 \pi R_{\rm A})$  comes naturally from the connection and it is defined with the size of apparent horizon,
$ R_{\rm A} =  1/\sqrt{H^2 + k/a^2}$.  Considering energy transfer flux passing through the apparent horizon with the first law of thermodynamics, one can achieve the Friedmann equation \cite{Cai:2005ra}. For $k=0$, the apparent horizon is just the Hubble length.  In this work, we will use the Hubble length scale as a flat geometry case of the $R_{\rm A}$ cutoff length scale.   A conclusive review on HDE models can be further studied in \cite{Wang:2016och}.

On the matter nature of HDE, the HDE could be vacuum energy or scalar field. This is such as quintom model which has a mixing of quintessential  kinetic  term and phantom kinetic term  (see e.g. \cite{CaiSaridakisSetare2010}). Instead of vacuum energy, the quintom field is taken as a holographic dark energy with phantom-crossing behavior. The model is known as hessence model \cite{Zhao:2007qy}.
Taking ability in unifying of inflation and phantom-crossing late time acceleration, a model was invented to accommodate the unifying picture. This  can be achieved either with the dilaton-like self-coupling of scalar kinetic term or with generalized version of the holographic cutoffs \cite{Nojiri:2005pu}.  The original HDE model when considered in Brans-Dicke gravity (Jordan frame),  using Hubble scale cutoff and particle horizon cutoff, fails to result in acceleration.  Only when using future event horizon cutoff, acceleration can be achieved \cite{Gong:2004fq}.  Another incorporation of scalar-tensor theories to the HDE model was reported in \cite{Bisabr:2008gu}.
In Brans-Dicke gravity, the HDE Hubble scale cutoff is viable only when scalar potential $V(\phi)$ is added to the theory  \cite{Liu:2009ha, Gong:2008br}.   Cosmology of non-minimal coupling to gravity (NMC) scalar field with original Hubble scale cutoff has been explored \cite{Ito2005, Setare:2008pc, Granda:2009zx}.  For the case of NMDC scalar field as HDE, de-Sitter solution has been studied at late time \cite{Kritpetch:2020vea}.  In this work, we consider the NMDC field in a flat universe filled with holographic vacuum energy and dust matter using Hubble scale cutoff. We study how the holographic term can alter inflationary parameters and we  examine  if  the HDE could rescue the NMDC inflation. Moreover we examine if NMDC field could help original HDE model with Hubble scale cutoff to achieve the acceleration that does not happen in Brans-Dicke gravity (as reported in \cite{Gong:2004fq}).   It is also interested that variation of gravitational strength could be used to constrain a range of the NMDC coupling in holographic dark energy model.
In section II, we introduce NMDC action and cosmological field equations. Slow-roll approximation and slow-roll parameters are derived in section III.
 In section IV, we give inflationary parameters and, for three  types of scalar potentials, we show parametric plots of the inflationary allowed range of the NMDC coupling and the field value. Scalar field phase portraits and acceleration regions are shown in section V. Constraint from variation of gravitational constant is analyzed in Section VI and we finally conclude this work in Section VII.

\section{Holographic vacuum energy and NMDC gravity effect }
The action of gravitational theory with scalar field non-minimal derivative coupled to gravity, dust matter and a cosmological constant  is given by
\bea
S &=&\int \d^4x\sqrt{-g}\left\{\frac{R}{16\pi G}-\frac{[\varepsilon g_{\mu\nu}+\kappa G_{\mu\nu}]}{2} (\nabla^{\mu}\phi) (\nabla^{\nu}\phi) - V   \right\}   +      S_{\rm m, \Lambda}\,.
\eea
The coupling constant $\k$ has $mass^{-2}$ dimension.
This action is a sub-class of Horndeski action as  $
G_2  =  -({\varepsilon}/{2}) g_{\mu\nu}(\nabla^{\mu} \phi)(\nabla^{\nu} \phi),\:  G_3 = 0, \: G_4 = (16 \pi G)^{-1},\:     G_5 = c_5 \phi = \phi {\k}/{2},  \: {\text{with}}   \: c_5 \equiv {\k}/{2}$     \cite{Horn, Deffayet:2011, Kobayashi}.
As a low-energy effective theory, there are modifications to the Einstein field equation,
\begin{equation}
	G_{\mu\nu} = 8\pi G \l( T^{(\rm {m})}_{\;\mu\nu}+T^{(\phi)}_{\;\mu\nu}+\kappa\Theta_{\mu\nu} \r)  -  \Lambda g_{\mu\nu},
\end{equation}
with the NMDC and free scalar conservation,
\begin{equation}
	[\varepsilon g^{\mu\nu}+\kappa G^{\mu\nu}]\nabla_{\mu}\nabla_{\nu}\phi =    V_{\phi},
\end{equation}
where $V_{\phi}\equiv \d V(\phi)/\rm{d}\phi,$  $T_{\mu\nu}^{(\rm m)}$ is a stress-energy tensor of dust matter field and
\begin{eqnarray}
	T^{(\phi)}_{\;\mu\nu} & = &  \ve(\nabla_{\mu}\phi) (\nabla_{\nu}\phi) - \frac{\ve}{2}g_{\mu\nu}(\nabla \phi)^2 - g_{\mu\nu}V(\phi), \\
	\Theta_{\mu\nu}  &=& -\frac{1}{2}(\nabla_{\mu}\phi)(\nabla_{\nu}\phi) R \,+\, 2 (\nabla_{\alpha}\phi) \nabla_{\left(\mu \right.}\phi R^{\alpha}_{\left.\;\nu\right)}  \,+\,  (\nabla^{\alpha}\phi) (\nabla^{\beta}\phi) R_{\mu\alpha\nu\beta}
	\,+\, (\nabla_{\mu}\nabla^{\alpha}\phi) (\nabla_{\nu}\nabla_{\alpha}\phi)  \no \\
	&& -\, (\nabla_{\mu}\nabla_{\nu}\phi) \Box\phi \,-\, \frac{1}{2}(\nabla\phi)^2 G_{\mu\nu}   \,+\,g_{\mu\nu}\bigg[-\frac{1}{2}(\nabla^{\alpha}\nabla^{\beta}\phi)(\nabla_{\alpha}\nabla_{\beta}\phi) + \frac{1}{2}(\Box\phi)^2  - (\nabla_{\alpha}\phi)(\nabla_{\beta}\phi) R^{\alpha\beta} \bigg].
\end{eqnarray}
Imposing Bianchi identity $\nabla^{\mu} G_{\mu\nu}=0$ and conservation of matter field $\nabla^{\mu}T^{({\rm m})}_{\;\mu\nu}=0$, hence the NMDC together with the free scalar field are conserved,
\begin{equation}
	\nabla^{\mu}[T^{(\phi)}_{\;\mu\nu}+\kappa \Theta_{\mu\nu}]  = 0.
\end{equation}
Considering that there is a holographic energy scale cutoff for the vacuum energy density $\rho_{\Lambda} = \Lambda/(8\pi G) =  {3c^2}/{(8\pi G L^2)}$, i.e. GR cosmic boundary is imposed as an IR quantum-gravitational modification of classical GR regime. The holographic effect is not described at classical action level but it is described phenomenologically in the Friedmann and other field equations.
In this situation, $\rho_{\Lambda}$ is not constant but scaling with cutoff length scale $L$ which is allowed to be time-dependent.
In spatially flat FLRW universe, derived from the Einstein field equation, the Friedmann equation reads,
\small
\begin{eqnarray}
H^2 &=&\frac{8\pi G}{3}\left[\frac{1}{2}\dot{\phi}^2(\varepsilon-9\kappa H^2)+V(\phi)+\rho_{\Lambda}+\rho_\m\right],     \label{F1}   \eea  or
\bea
H^2 & = & \frac{8\pi G_{\text{eff}}}{3}\left[\frac{\ve}{2}\dot{\phi}^2+V(\phi)+\rho_{\Lambda}+\rho_\m \right],   \label{F2}
\end{eqnarray}
\normalsize
where  $\ve= \pm 1$ for canonical and phantom case and $\kappa$ is the NMDC coupling constant. The scalar potential is $V(\phi)=V_0 \phi^n$ and $V = V_0 \exp{(-\beta \phi)}$. Densities
 $\rho_{\Lambda}, \rho_{\rm m}$ are the holographic vacuum energy density and the dust matter density respectively. The energy density of holographic vacuum energy and dust matter are conserved separately as
  $\dot{\rho}_{\Lambda} +3H(\rho_{\Lambda} + P_{\Lambda})=0, \dot{\rho}_{\rm m} = -3H\rho_{\rm m}$ where $P_{\Lambda} $ is pressure of holographic vacuum energy. The modified Friedmann equation can be represented  in terms of modification of the kinetic term of the scalar field namely $  \dot{\phi}^2_{\text{eff}} =    (\dot{\phi} \sqrt{\varepsilon-9\kappa H^2})^2  $  (in equation (\ref{F1})) in which the kinetic energy contribution is  enhanced for $\k < 0$  and is reduced for $\k > 0$.
  The other way of consideration is to view the NMDC effect as a modification of the gravitational constant $G$ (in equation (\ref{F2})),
 \begin{eqnarray} \label{5}
	G_{\text{eff} }(\dot{\phi}) \; \equiv \;  \frac{G}{1+12\pi G \k \dot{\phi}^2}\,,
\end{eqnarray}
 where factorization of $G_{\text{eff}}$ does not affect  $V(\phi), \rho_{\Lambda}$ and $\rho_\m$.  For $\k > 0$, $G_\text{eff}$  is less than the usual  $G$  and for $\k < 0$ the result is opposite.
 The Klein-Gordon equation describing conservation of NMDC scalar field energy density can be viewed as
\begin{equation}\label{kg}
	\ddot{\phi}+3H \dot{\phi}   \l( 1  -     \f{2 \k \dot{H} }{\varepsilon - 3\kappa H^2}    \r)
=-\frac{V_\phi}{\varepsilon - 3\kappa H^2}\,,
\end{equation}
where the NMDC modification appears in the damping and potential terms.
Holographic vacuum energy density does not interact with the NMDC field hence there is no explicit holographic effect in the Klein-Gordon equation. However, implicitly holographic effect contributes via $H$ and $\dot{H}$.   We use apparent horizon cutoff
$L = R_{\rm A} =  1/\sqrt{H^2 + k/a^2} = H^{-1}$  in flat universe. There are two non-equivalent ways in considering the system:
\begin{enumerate}
  \item  The Friedmann equation (\ref{F1}) can be viewed as a description of a standard flat FLRW universe filled with NMDC  field and other matters. In this case, the holographic energy density should be defined as
 \begin{equation}\label{rhol0}
 	\rho_{\Lambda}\;   \equiv \;\frac{3c^2H^2}{8\pi G}\,.
 \end{equation}
  \item  The Friedmann  equation (\ref{F2}) can be viewed as a description of FLRW universe with time-dependent gravitational coupling, $G_{\text{eff}}$. The universe is filled with canonical (phantom) scalar field and other matters.  In this cosmological consideration, we have
 \begin{equation}\label{rhol}
 	\rho_{\Lambda}\;  \equiv   \;\frac{3c^2H^2}{8\pi G_{\text{eff}}}\;=\;\frac{3c^2}{8 \pi G}(1+12\pi \kappa G \dot{\phi}^2)H^2\,,
 \end{equation}
 as it is not only a function of $H$ but also of $\dot{\phi}$.  In this case, the gravitational constant can be viewed as varying  gravitational strength parameter, $G_\text{eff}(\dot{\phi})$.
\end{enumerate}
 We shall follow the second consideration in this work.  Introducing of the NMDC field  hence prevents the problematic dust-like solution in \cite{Hsu} in which  $c$  is set to $1$ and it is without scalar field component.
 Further factorization, the Friedmann equation (\ref{F2}) can be expressed in form of
\begin{eqnarray}\label{h2l}
	H^2 = \f{8 \pi \tilde{G}_{\text{eff}}}{3}\l[\frac{\ve}{2}\dot{\phi}^2+V(\phi)+\rho_\m\r],   \label{F3}
\end{eqnarray}
where the new effective gravitational constant,
\be
\tilde{G}_{\text{eff}}  =  \f{G}{(1-c^2)(1+12 \pi G \k \dot\phi^2)}\,,
\ee
includes both NMDC and holographic contributions. This can be interpreted as there are only dust and a canonical (phantom) scalar field
in flat FLRW universe with a new modified gravitational coupling $\tilde{G}_{\text{eff}}$. The second field equation is\footnote{The equation (\ref{dotH}) is obtained by doing time derivative of equation (\ref{F3}), using $\ddot{\phi}$ of the Klein-Gordon equation in the result and finally factorizing of $\dot{H}$. This is unlike the procedure in \cite{Kritpetch:2020vea} where $P_{\Lambda} = -\rho_{\Lambda}$ is assumed.}
\bea
\dot{H}&=& \f{4 \pi \tilde{G}_{\text{eff}} }{A}\Bigg\{  -\varepsilon \dot{\phi}^2
             + 9 \k H^2(1-c^2) \bigg[ \dot{\phi}^2 + \f{V_{\phi} \dot{\phi}(2-3c^2) }{9 H (\varepsilon - 3 \k H^2 )(1-c^2)} \bigg] - \rho_{\rm m}  \Bigg\}\,,    \label{dotH}
\eea
where $A  \equiv  1 -  8 \pi \tilde{G}_{\text{eff}} \dot\phi^2  \l\{  \l[ \k \ve - 9 \k^2 H^2 (1-c^2) \r]/(\ve -3\k H^2)  \r\}  \,. $

\section{Slow-roll approximation}
In the slowly rolling regime with negligible dust density, one can approximate that $\dot{\phi}^2 \ll V $
%and $\k \dot{\phi}^2 \ll V $,
hence the approximated Friedmann equation is,
\begin{equation} \label{simh}
H^2  \;\simeq\; \f{V}{3 M^2_\p (1-c^2)(1+12\pi G \k \dot\phi^2)}\,.
\end{equation}
With sub-Planckian value, i.e.  $|\k| < 1$ (in the unit of $M_{\p} = 1$), $\k^2 \ll 1$ and with the slow-roll approximation $\dot{\phi}^2 \ll V $, hence the factor $A \approx 1$\,.  In the slow-roll regime, the field acceleration is negligible, $\ddot{\phi} \approx 0$, therefore the Klein-Gordon equation is
\be
\dot{\phi}   \;\simeq\;   \f{- V_{\phi}}{3 H (\ve - 3 \k H^2 - 2 \k \dot{H})}\,.  \label{KGslow}
\ee
%%%%%%%%%%%%%%%%%%%%%%%%%%%% here
We express slow-roll parameters,
\be
\e  \equiv -\f{\dot{H}}{H^2}\,,\;\;\;\;
  \delta    \equiv  -\f{\ddot{\phi}}{H \dot{\phi}}\,,\;\;\;\;
\eta  \equiv \f{\dot{\e}}{H \e}\,,\;\;\;\;
 \text{and}\;\;\;\; \eta_{V}    \equiv  M^2_\p \l| \f{V_{\phi \phi}}{V}\r|\,.
\ee
From the equation (\ref{simh}) with equation (\ref{KGslow}), the $\dot{H}$ equation is
\be
\dot{H} \;\, \simeq\;\, \frac{V_{\phi} \dot{\phi}}{6 M^2_\p H (1-c^2)(1+12\pi G \k \dot\phi^2)} \;\,   \simeq \;\,  \f{- V_{\phi}^2}{6 V (\ve - 3 \k H^2 - 2 \k \dot{H})}\,.  \label{dotH1}
\ee
Hence, the first slow-roll parameter is
\be
\e \:\simeq\:  \e_V \,\l[\f{(1-c^2)(1+12\pi G \k \dot\phi^2)}{\ve - 3 \k H^2 - 2 \k \dot{H}}\r]\,,
\ee
where $\e_V \equiv  \l({M^2_\p}/{2}\r) \l({V_{\phi}}/{V}\r)^2  $.
In the slow-rolling regime,
\bea
 \dot{\phi}^2 & \ll & (12\pi G \k)^{-1}\,, \label{approx1}  \\ \text{and}\;\;\;\;\;\;
 |\ddot{H}| &\ll& |H \dot{H}| \;\ll\; |H^3|\,, \label{approx2}
 \eea
implying $\e \ll 1$ and $ 3 \k H^2 + 2 \k \dot{H} = 3 \k H^2[1 - (2/3)\e] \simeq   3 \k H^2\,.$ Therefore the Friedmann equation (\ref{simh}) and the equation (\ref{dotH1}) are approximated to
\be
H^2 \,\simeq\, \f{V}{3 M^2_\p (1-c^2)}\,,\;
\;\; \dot{H}  \, \simeq \,   \f{- V_{\phi}^2}{6 V \l(\ve - \f{\k V}{{M^2_\p}(1-c^2)} \r)}\,.   \label{HanddotH}
\ee
Hence,
\bea
\e \;&\simeq& \; \e_V  \l[\f{(1-c^2)}{\ve - 3 \k H^2}\r]\,\; \simeq \,\;
   \e_V \l[ \f{(1-c^2)}{ \ve - \f{\k V}{M^2_\p (1-c^2)}}\r]      \,.
\eea
Using slow-roll approximations with (\ref{approx1}), (\ref{approx2}) and (\ref{HanddotH}), the second and the third slow-roll parameters are
\bea
\delta \;&\simeq &\;  \eta_{V}    \l[ \f{(1-c^2)}{ \ve - \f{\k V}{M^2_\p (1-c^2)}}\r]      \:-\:  \e  \: +\:
 \f{\k   \l(V_{\phi}^2/V\r) }{\l(\ve - \f{\k V}{M^2_\p (1-c^2)}      \r)^2}   \,, \\
 \eta  \;& \simeq &\;  -2 \eta_{V} \l[ \f{(1-c^2)}{ \ve - \f{\k V}{M^2_\p (1-c^2)}}\r]   \: +\:
   4 \e \: - \:
 \f{\k  \l(V_{\phi}^2/V\r) }{\l(\ve - \f{\k V}{M^2_\p (1-c^2)}      \r)^2}   \,.
 \eea
This reduces to standard canonical scalar field in GR without holographic effect when $\ve = 1, c = 0$ and $\k = 0$. These are $\delta \,\simeq\, \eta_V - \e,\;\: \eta \,\simeq\, -2 \eta_V + 4\e \:$   or  $\: \eta \,\simeq\,  2 \eta_V - 4\delta$.

\section{Inflationary parameters}
In general, spectral index $n_{\text{s}}$ is related to slow-roll parameters as
$n_{\text{s}} -1 \;\simeq \; -2\e - \eta \; =\; - 4 \e + 2 \delta $. For our model, this is
\bea
n_{\text{s}} -1
&\simeq & -4 \l[\f{(1-c^2) \e_V}{\ve - \f{\k V}{M^2_\p (1-c^2)} } \r] \:+\: 2 \delta \: -\:   \f{\k}{\l(\ve - \f{\k V}{M^2_\p (1-c^2)} \r)^2} \f{V_{\phi}^2}{V}\,, \no \\
&\simeq &  \l[\f{(1-c^2)}{\ve - \f{\k V}{M^2_\p (1-c^2)} } \r]\l(-6\e_V + 2 \eta_V \r)   \:+\:    \f{\k}{\l(\ve - \f{\k V}{M^2_\p (1-c^2)} \r)^2} \f{V_{\phi}^2}{V}\,.
\eea
Tensor-to-scalar ratio is
\be
r\, =\, 16 \e\,  \simeq \, 16 \l[\f{(1-c^2) }{\ve - \f{\k V}{M^2_\p (1-c^2)} } \r] \e_V\,. \label{r}
\ee
Planck 2018 CMB data \cite{Planck:2018vyg} gives a constraint, $n_{\text{s}} = 0.965 \pm  0.004$  for single field inflationary model. When
combining with BICEP-Keck 2015 data on B-mode polarization, the tensor-to-scalar ratio is given an upper limit of $r_{0.002} < 0.06$ at 95 \% CL.
From now on, we consider canonical scalar field case with $\ve = 1$ and positive potential $V(\phi) > 0$.  For brevity, we define $\alpha(\k, c) \equiv  \ve - {\k V}/[M^2_\p (1-c^2)]$.
Using equation (\ref{r}) with the constraint, $r_{0.002} < 0.06$,
the denominator $\alpha$ is considered in two cases, i.e. $\alpha > 0$ and $\alpha < 0$. For $\alpha > 0$, we have $\k < \ve M^2_\p (1-c^2)/V$ and the opposite is $\alpha < 0$, which gives $\k > \ve M^2_\p (1-c^2)/V$.
The e-folding number for this model is
\be
N \,=\, \int_{t_{\rm i}}^{t_{\rm f}} H \d t
   \,\simeq \,  \int^{\phi_{\rm i}}_{\phi_{\rm f}} \l[ \f{\ve - \f{\k V}{M^2_\p (1-c^2)} }{1-c^2}\r] \f{1}{\sqrt{2 \e_V}} \f{\d \phi}{M_\p}\,,
\ee
where $\phi_{\rm f}$ is set to $0$ and from now on, we rename $\phi_{\rm i}$ to $\phi$. We shall consider these parameters for a specific type of potential.

\subsection{Inflation with power-law potential}

Inflationary potential is chosen to be power-law type, $V = V_0 \phi^n $ with $n = 2, 4$ and $V_0 \equiv \lambda M_\p^{4-n}$. In case of $n=2$, $\lambda \equiv (1/2)m^2 M_\p^{-2}$ and in case of $n=4$, $\lambda \equiv 1/4$.  Shortcoming of non-holographic NMDC inflation with power-law potential is re-stressed recently in \cite{Avdeev:2022ilo} using $n_{\text{s}}$ and $r$  parameters. Let us see if the holographic effect given to the NMDC model would make any difference. The e-folding number in this case is
\be
N  =  N_{\text{GR}} \l[\f{\ve}{1-c^2} -  \f{2\k}{M_\p^2(1-c^2)^2} \l( \f{V_0 \phi^n}{n + 2} \r)     \r]\,,     \label{yak}
\ee
where $ N_{\text{GR}} \equiv \phi^2/(2 n M_\p^2) $. The value $N \geq 60$ is needed to solve the horizon problem.
%%%

In case of $n = 2$ with $N = 60$, the equation (\ref{yak}) gives
\be
\phi^2 = \f{M_\p^2(1-c^2)}{\k V_0} \l( \ve \pm \sqrt{1 - 480 \k V_0}\r)\,.   \label{phi_n2}
\ee
  For $\alpha > 0$, we have  $\k < \ve M^2_\p (1-c^2)/(V_0 \phi^n)$. The upper bound $r < 0.06$ results in a bound on $\k$ value,
\be
\k\,<\,\f{\ve M_\p^2 (1-c^2)}{V_0\phi^n}  -  \f{ M_\p^4 n^2 (1-c^2)^2}{V_0\phi^{n+2}}\f{8}{0.06}\,.  \label{kboundN}
\ee
In this case, using equation (\ref{phi_n2}) in equation (\ref{kboundN}), we found two possibilities, $\k > 0$ or $\k < 0$. For $\k > 0$, valid range of the NMDC coupling is
$ 0 < \k V_0 \leq 1/480\, $
or $0 < \k V_0 \leq 0.002$. For $\k < 0 $ case, positive and negative roots of equation (\ref{phi_n2}) gives
$
-0.206 < \k V_0 < -0.019 $ and $
-0.019 < \k V_0 < 0\,
$
respectively. In combination, for $\alpha > 0$ case, the allowed range of $\k$ is approximately
$
-0.206  \,<\, \k V_0 \, \leq  0.002 \,.
$
For  $\alpha < 0$,  only $0 <  \ve M^2_\p (1-c^2)/(V_0 \phi^n)  < \k $ is allowed and the valid range of the NMDC coupling is
$ 0 < \k V_0\,. $
When $n=4$, the equation (\ref{yak}) is not simple.
We perform parametric plots of $\k$ and $\phi$  in figure \ref{fig1} for these power-law potentials with other constraints, e.g. power spectrum index and number of e-foldings required to solve the horizon problem ($N \geq 60$). As seen in the figure, $n_{\text{s}}$ and $r$ viable regions do not overlap each other. Both partially overlap with only the $N \geq 60$ region.

\subsection{Inflation with exponential potential}
\begin{figure}[t]  \begin{center}
\includegraphics[width=19.0cm,height=12.2cm,angle=0]{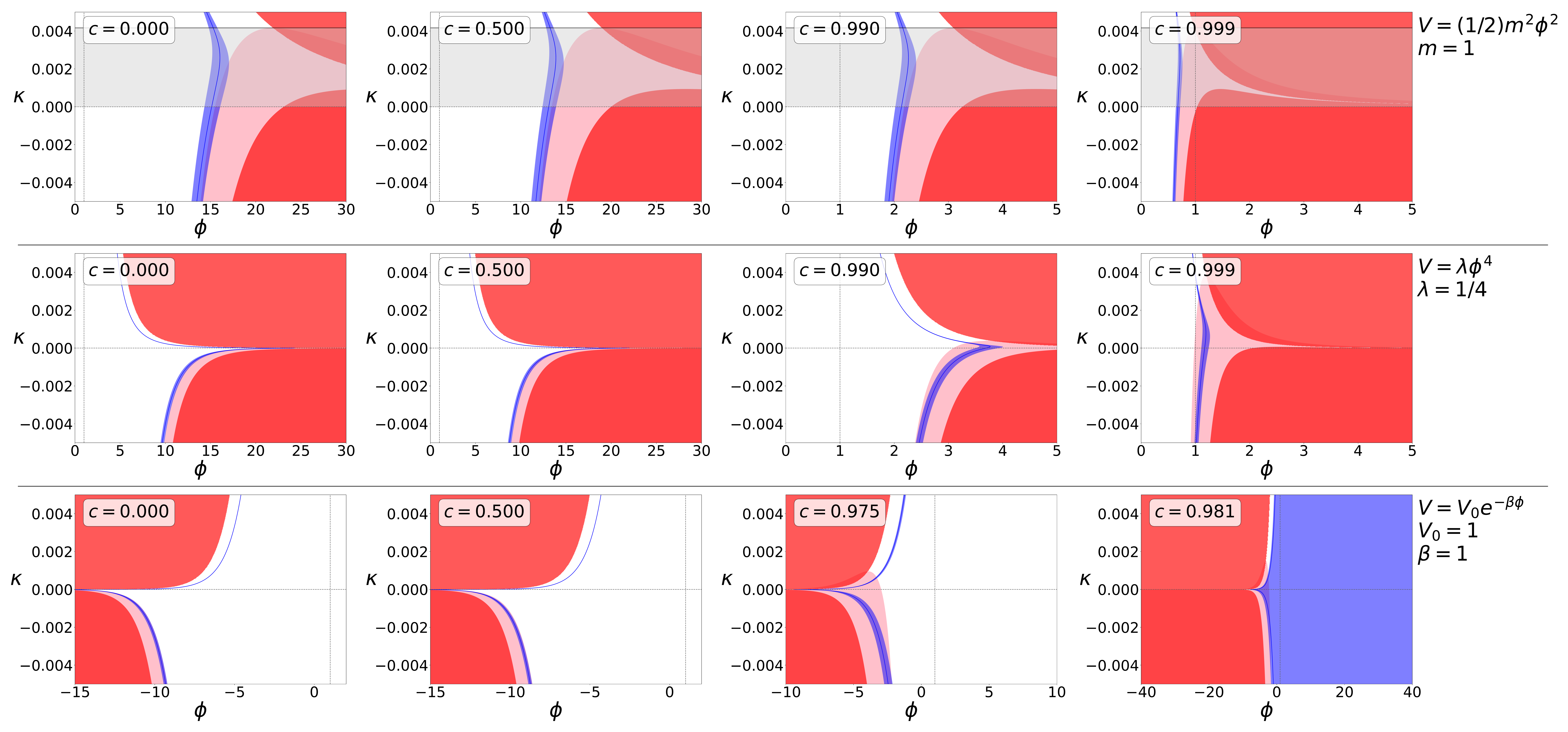}  \end{center}
\caption{Parametric plots of $\k$ and $\phi$ with the constraints $ n_{\text{s}} = 0.965 \pm  0.004$, $r_{0.002} < 0.06$ \cite{Planck:2018vyg} and the requirement $N \geq 60$.
The blue shade represents constraint from $n_{\text{s}}$ (mean and error bar). Allowed region of $r$ with the upper limit of $0.06$ is in red.  Pink shade represents region with $N \geq 60$.
Top, middle and bottom rows are of the case $V = (1/2)m^2\phi^2, V = \lambda \phi^4 $ and $V = V_0 \exp(-\beta \phi)$ accordingly.
In the top row, the grey shade ranges from $0$  to a value less than $1/240 \sim 0.004$. It is requirement of non-imaginary value of the term $\l( \ve \pm \sqrt{1 - 480 \k V_0}\r)$  in the equation (\ref{phi_n2}) when assuming positive $\k$.
\label{fig1}}   \end{figure}

Considering exponential potential, $V = V_0 \exp{(-\beta \phi)}$, upper bound on $r$ results in
\be
\k\,<\,   \f{e^{\beta \phi}}{V_0} \l[  \ve M_\p^2(1-c^2)  - \l(\f{8}{0.06}\r) \beta^2 (1-c^2)^2 M_\p^4 \r] \,. \label{kboundfour}
\ee
Parametric plot $\k$ versus $\phi$ of this case is shown in the bottom row of figure \ref{fig1}. In the exponential potential case, one can also see that both $n_{\text{s}}$ and $r$ viable regions do not overlap
but they partially overlap with region of $N \geq 60$.

\begin{figure}[t]  \begin{center}
\includegraphics[width=19.0cm,height=12.2cm,angle=0]{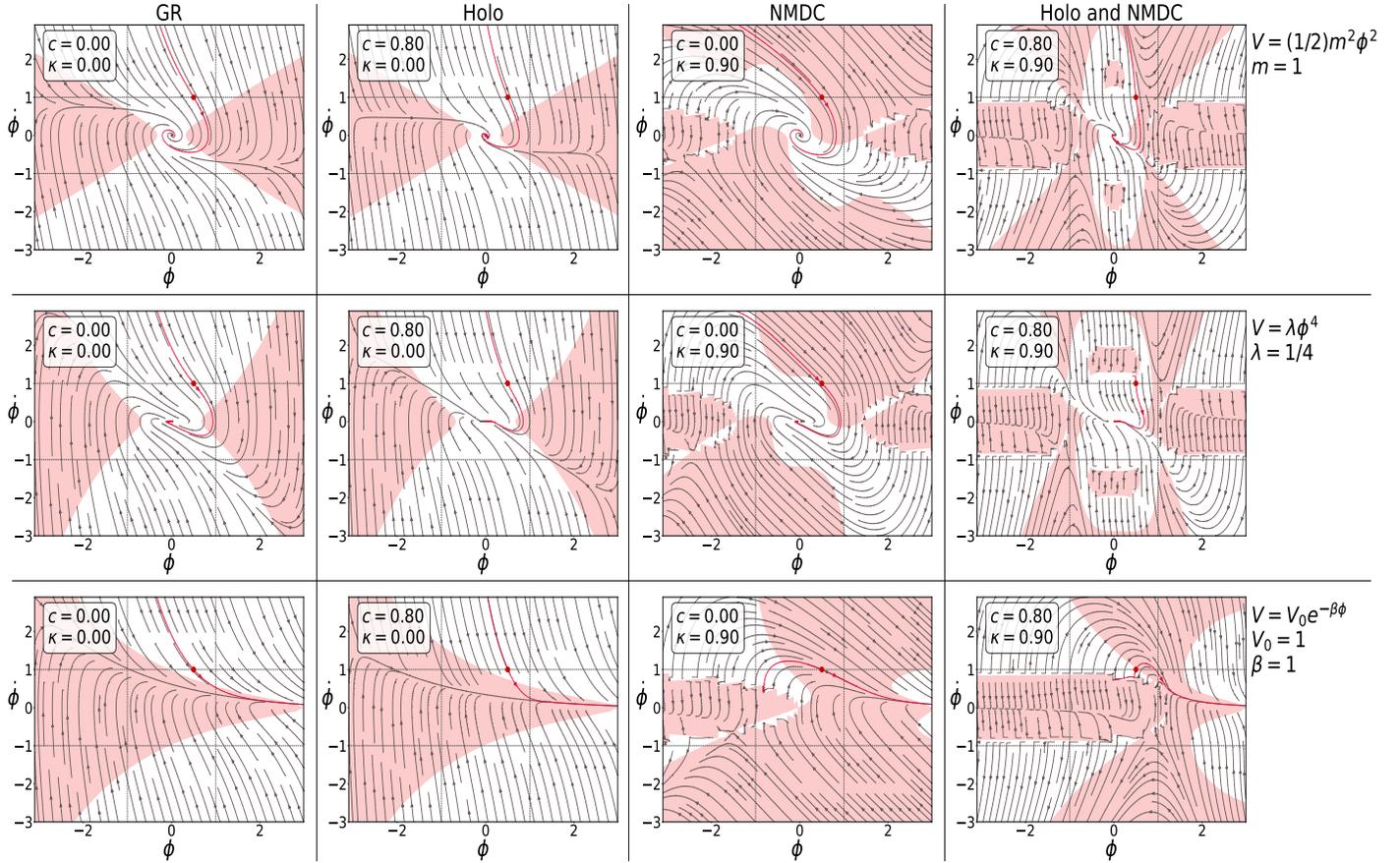}  \end{center}
\caption{Phase portraits, on the plane $\rho_{\rm m} = 0.1$, represent phase paths and acceleration regions (pink shade). Each row is of each type of potentials labeled on the right of the figures. Each column is of the GR  ($c = 0$  and $\k = 0$), holographic ($c \neq 0$  and $\k = 0$),  NMDC ($c = 0$  and $\k \neq 0$) and mixed NMDC holographic ($c \neq 0$  and $\k \neq 0$) cases. Red dot in each phase portrait locates at $\dot{\phi} = 1$  and $\phi = 0.5$.   \label{fig2}}   \end{figure}

\section{Phase portraits}
%%%%%%%%%%%%%%%%%%%%%%%%%%%% here
We investigate the holographic NMDC model as dark energy in the late universe.
With the acceleration condition, $\ddot{a}/a\, = \, \dot{H}  + H^2  \,>\, 0 $, acceleration regions are presented with pink shade in  figure \ref{fig2}.  Evolution of the system with cosmic time can be presented as phase portraits. The closed autonomous system of $\{\phi, \dot{\phi}, H, \rho_\m   \}$ is\
\begin{align*}
\dot{\phi} &\equiv \psi, \qquad \dot{\rho}_{\rm m} = -3H\rho_{\rm m} , \\
\dot{H} &= \frac{4\pi G}{(1-c^2)(1+12 \pi G \kappa \psi^2) A }\Bigg\{-\varepsilon \psi^2
+ 9\kappa H^2(1-c^2)\left[\psi^2 +\frac{V_\phi\psi (2-3c^2) }{9H(\varepsilon - 3\kappa H^2)(1-c^2)}
 \right] -\rho_{\rm m} \Bigg\},
 \\
 \dot{\psi}&=-3H \psi-\frac{V_\phi}{\varepsilon - 3\kappa H^2} + \frac{24 \pi G \kappa H \psi \Bigg\{-\varepsilon \psi^2
+ 9\kappa H^2(1-c^2)\left[\psi^2 +\frac{V_\phi\psi (2-3c^2) }{9H(\varepsilon - 3\kappa H^2)(1-c^2)}
 \right] -\rho_{\rm m} \Bigg\}  }{(\varepsilon - 3\kappa H^2)(1-c^2)(1+12 \pi G \kappa \psi^2) A},
\end{align*}
 The autonomous system can be expressed without any needs of approximation.  The phase portraits are shown in figure \ref{fig2}.  Using approximations $\ddot\phi \approx 0, \dot\phi^2 \ll V, \dot{H} \ll H^2$. Late time trajectory is approximated,
\begin{eqnarray}
\dot{\phi} &\simeq& \frac{-V_\phi}{\frac{\sqrt{3V}}{M_{\rm p}\sqrt{(1-c^2)}} \left(\varepsilon - \frac{\kappa V}{M^2_{\rm p} (1-c^2)} \right)}.  \label{latetime}
\end{eqnarray}
From figure \ref{fig2},   Late time behaviors of the system are considered as the behavior in the present epoch.  If the scalar field is canonical (phantom), $\k = 0$, holographic term does not affect acceleration region, i.e. acceleration conditions of the GR and holographic cases are the same.
In third and forth columns of the figure, NMDC term can affect the acceleration regions. The holographic term can alter the acceleration regions only in presence of NMDC field. For all three types of potential, we see qualitatively that the NMDC effect results in anatomy of the dynamics, i.e. it results in new late-time attractor trajectories in the acceleration regions, and that the holographic effect only changes shape of the acceleration regions and pattern of the phase paths.  The holographic effect does not alter anatomy of the phase portraits.
NMDC dynamical study  in \cite{Skugoreva:2013ooa} has been generalized in \cite{Matsumoto:2017gnx} where qualitative analysis of NMDC gravity was reported in detail. In   \cite{Matsumoto:2017gnx}, asymptotes of the dynamical system are found and the correponding asymptotic Hubble rates are reported therein. 
Holographic effect modifies all the asymptotic Hubble rates of  NMDC gravity case via  the gravitational constant $G$ and via the NMDC coupling $\kappa$. That is to say, the holographic effect changes $G$ to $G/(1-c^2)$ and $\kappa$ to $\kappa (1-c^2)$.  For instance in \cite{Matsumoto:2017gnx}, two late time asymtotic Hubble rates of the NMDC gravity case are $H = \sqrt{8 \pi G V/{3}}$ and $H = 1/\sqrt{3\kappa}$. The holographic effect modifies these to $H = \sqrt{8 \pi G V/{3(1-c^2)}}$ and   $H = 1/\sqrt{3\kappa (1-c^2)}$. Since the modification term, $(1-c^2)$ 
is only a constant, it does not change any anatomy of the NMDC phase portraits but it only changes pattern scale of the NMDC phase portraits as seen in  figure \ref{fig2}.

\section{Variation of gravitational constant}
Variation of gravitational constant can be constrained by several observations.
For instance, data from gravitational-wave standard sirens and supernovae provide $\dot{G}/G|_{t_0} \lesssim 3\times10^{-12} \,  \text{year}^{-1}$ \cite{Zhao:2018gwk} while constraints in the same order of magnitude are confirmed by observations of pulsars \cite{Kaspi1994, Zhu:2018etc}, lunar laser ranging \cite{Williams2004} and BBN \cite{Copi2004, Cyburt2004}.
Recent observations such as gravitational wave observation of binary neutron stars and local gravitational acceleration measurement put weaker constraints on the variation. The binary neutron stars GW170817 provides a bound of
$   -7\times  10^{-9} \,  \text{year}^{-1} \lesssim  \dot{G}/G|_{t_0}  \lesssim 5\times 10^{-8}\,  \text{year}^{-1}$  at present \cite{Vijaykumar2021} while local gravitational acceleration measurement gives  $\dot{G}/G|_{t_0}  < 5.61 \times 10^{-10}\,  \text{year}^{-1}$  \cite{Dai:2021jnl}.  Here, variation of gravitational constant in NMDC holographic gravity is given by
\be
\frac{\dot{\tilde{G}}_{\text{eff}}}{\tilde{G}_{\text{eff}} }   \,=\,   \frac{\dot{G}_{\text{eff}}}{G_{\text{eff}} } \,=\,  \frac{-24\pi G\kappa\dot{\phi}\ddot{\phi}}{1+12\pi G\kappa\dot{\phi}^2}      \,   \simeq  \,      \f{-3 \kappa}{M^2_{\p}}\dot{\phi}\ddot{\phi}\,.     \label{Gdot}
\ee
The holographic term $ 1-c^2 $  does not affect the variation of gravitational constant explicitly. However, implicitly there are holographic effects from $ c^2 $ terms in the equation (\ref{dotH}) which is used in the Klein-Gordon equation (\ref{kg}) to find the $\ddot{\phi} $  term in the equation   (\ref{Gdot}). With the upper bound
$ \dot{G}_{\text{eff}}/G_{\text{eff}}|_{t_0} \lesssim 3 \times 10^{-12} \,  \text{year}^{-1}$
given by \cite{Zhao:2018gwk}, the variation (\ref{Gdot}) reads
\be
10^{-12} \,  \text{year}^{-1}   \,  \gtrsim  \,      \f{- \kappa}{M^2_{\p}} \dot{\phi}\ddot{\phi}\,.    \label{hhhk}
\ee
Considering late-time trajectory (\ref{latetime}), for both power-law and decreasing exponential potentials, at late time $V (\phi) \ll  1$, hence the $\k V$  term is negligible (for $\ve = 1$). We can finally solve the equation (\ref{latetime}) to obtain  late-time scalar field exact solutions for each type of potentials.

\subsection{Variation of gravitational constant: Power-law potentials}

\subsubsection{The case of  $V = (1/2)m^2 \phi^2$  potential}
For $V = (1/2)m^2 \phi^2$, the late-time approximated solution is
\be
\phi(t) \:    \simeq   \:  - C  \,  (t - t_0) + \phi_0 \,,
\ee
where $C  \equiv  \sqrt{[{2 (1-c^2)}/{3}]} \,  m  M_{\p}$.   This  gives $\dot{\phi}  \simeq  - C$  and $ \ddot{\phi}  \simeq  0 $.    In order to avoid this null result, instead we approximate the Klein-Gordon equation (\ref{kg}) as\footnote{Late time is considered as the present time. In the Klein-Gordon equation (\ref{kg}),  the terms $\k H^2$ and $\k H \dot{H} \dot{\phi}$ are negligible because $\dot{\phi} \ll  1$  and $H$  is very small, i.e.  $H_0 \simeq 68 \; \text{km}/(\text{sec}\cdot \text{Mpc})  \simeq  22 \times 10^{-19} \;\text{sec}^{-1}$.},   $\ddot{\phi} \simeq  - 3 H \dot{\phi} - V_{\phi}$. The relation (\ref{hhhk}) is hence
\be
10^{-12} \,  \text{year}^{-1}   \:\,  \lesssim  \:\,      \f{ \kappa}{M^2_{\p}}  C \l[ 3 H_0 C   - m^2 \phi(t_0)  \r] \,.
\ee
At present, $t = t_0$  and $\phi(t_0) = \phi_0 \simeq 0$.    With $H_0\simeq 68 \; \text{km}/(\text{sec}\cdot \text{Mpc}) $ and  $M_{\p} = 1$,  we can constrain $\k$  for a particular value of $c$ and of mass, $m$ of the scalar field. The result is presented in table \ref{TableG}\,.

\subsubsection{The case of $V = \lambda \phi^4$  potential}
For $V = \lambda \phi^4$ , the late-time approximated solution is
\be
\phi(t) \:       \simeq   \:    \phi_0 \exp{\l\{-\l[ M_{\p}\sqrt{\f{(1-c^2)}{3\lambda}}\r]\,(t - t_0) \r\}}   \,.
\ee
Using this solution in the relation (\ref{hhhk}), considering that at present, $t = t_0$  and $\phi(t_0) = \phi_0 \simeq 1$, we obtain
\be
10^{-12} \,  \text{year}^{-1}   \:\,  \lesssim  \:\,      { \kappa}{M_{\p}}   \phi_0^2 \l(  \sqrt{\f{1-c^2}{3 \lambda}}        \r)^3               \,.
\ee
When setting $\phi_0 = 1, M_{\p} = 1$  and $\lambda =  1/4$, the constraint on $\k$ is shown in  table \ref{TableG}\,.

\subsection{Variation of gravitational constant: $V = V_0 \exp{(-\beta \phi)}$  case}
For $V = V_0 \exp{(-\beta \phi)}$ , the late-time approximated solution is
\begin{equation}
\phi(t) \:       \simeq   \:    \f{2}{\beta}   \ln{\l[ \f{\beta^2}{2}   M_{\p} \sqrt{\f{V_0 (1 - c^2)}{3} } (t-t_0)  \,+\,   \exp{(\beta \phi_0 / 2)}    \r]}  \,.
\end{equation}
The relation (\ref{hhhk}), at present, $t = t_0$  and $\phi_0 =  0$, is
\be
10^{-12} \,  \text{year}^{-1}   \:\,  \lesssim  \:\,      { \kappa}   \f{\beta^3}{6} V_0 (1-c^2) \exp{(-\beta \phi)}        \,.
\ee
Setting $V_0 = 1, M_{\p} = 1$  and $\beta =  0.5, 1$,  the constraint on $\k$ is shown in  table \ref{TableG}\,.

\begin{table}[!h]
\begin{tabular}{|c|c|c|c|c|}
\hline
$\;\:\;\:c\;\:\;\:$ &   $ \;\: \;\: V = (1/2) m^2 \phi^2 \;\: \;\: $  & $\;\: \;\:\;\:\;V = (1/4) \phi^4 \;\;\: \;\:\;\: $ & $\; V = V_0 \exp{(-\beta \phi)},  \beta = 1\; $   &  $\;\: V = V_0 \exp{(-\beta \phi)},  \beta = 0.5 \;\:$  \\
\hline \hline
0     &   $0.0072/m^2  \lesssim  \k $    &  $2.1 \times 10^{-20} \lesssim  \k $   & $ 5.2 \times 10^{-19} \lesssim  \k$  & $2.5  \times 10^{-18}  \lesssim  \k $    \\
 0.3  &  $0.0079/m^2  \lesssim  \k $   &  $ 2.4 \times 10^{-20}  \lesssim  \k $  &  $  5.7 \times 10^{-19} \lesssim  \k $   &  $2.8  \times 10^{-18}  \lesssim  \k $  \\
 0.5  &  $0.0096/m^2  \lesssim  \k $   &  $ 3.2 \times 10^{-20}  \lesssim  \k $  &  $  6.9 \times 10^{-19} \lesssim  \k $   &  $3.3  \times 10^{-18}  \lesssim  \k $  \\
  0.7  &  $0.014/m^2  \lesssim  \k $   &  $ 5.7 \times 10^{-20}  \lesssim  \k $  &  $  1.0 \times 10^{-18} \lesssim  \k $   &  $4.9  \times 10^{-18}  \lesssim  \k $  \\
0.8  &   $0.02/m^2 \lesssim  \k   $    &  $ 9.5 \times 10^{-20}  \lesssim  \k $  &  $ 1.4 \times 10^{-18}  \lesssim  \k $ & $ 7.0  \times 10^{-18}  \lesssim  \k $ \\
0.9  &   $0.038/m^2 \lesssim  \k $   &  $2.5 \times 10^{-19} \lesssim  \k $  & $ 2.7  \times 10^{-18}  \lesssim  \k$  & $  1.3  \times 10^{-17}  \lesssim  \k $ \\
\hline
\hline
\end{tabular}
\caption{Variation of gravitational constant gives constraint on $\k$  for a range of constant $c$ using three types of scalar potential.  The constraint is derived from the upper bound, $\dot{G}/G|_{t_0} \lesssim 3\times10^{-12} \,  \text{year}^{-1}$   obtained from gravitational-wave standard sirens and supernovae data \cite{Zhao:2018gwk}. }
\label{TableG}
\end{table}

\section{Conclusions}
In this work we consider non-minimal derivative coupling (NMDC) to gravity term  in flat FLRW universe with the holographic effect from Bekenstein-Hawking entropy. The Hubble horizon is used as holographic cutoff scale. This cutoff is the flat geometry case of the apparent horizon cutoff.  With the Hubble horizon cutoff, the holographic vacuum energy density is  $\rho_{\Lambda}  =  3c^2 H^2/(8 \pi G)$ where we consider $0 < c <1$ in this work. The case $c = 1$ is a separatrix with much distinct dynamical behaviors and it is worth considered in future.  The NMDC effect can be considered as either modification in the kinetic scalar term or modification in the gravitational constant term. In the Friedmann equation (\ref{F2}), NMDC modification to the gravitational constant results in effectively time-varying behavior, leaving the scalar field kinetic density in canonical (phantom) form. The NMDC effect is cosmological hence the gravitational constant in definition of holographic vacuum energy density should include the NMDC effect,  that is $\rho_{\Lambda}  =  3c^2 H^2/(8 \pi G_\text{eff}(\dot{\phi}))$ where
$G_{\text{eff} }(\dot{\phi})  \equiv   {G}/({1+12\pi G \k \dot{\phi}^2})$.
 Considering $0 < c < 1$, in the Friedmann equation, the holographic effect can be factorized to include in $G_\text{eff}$, making a new effective gravitational constant, $\tilde{G}_{\text{eff}}  \equiv   {G}/[{(1-c^2)(1+12 \pi G \k \dot\phi^2)}]$. One can see that the NMDC term  reduces the effective strength of gravitational constant for $\k > 0$ (and the opposite effect for $\k < 0$) while the holographic effect ($c^2$ term) enhances the gravitational strength.  Using new effective gravitational constant $\tilde{G}_{\text{eff}}$, matter components in the Friedmann equation are only  canonical (phantom) scalar field and the barotropic matter (as in equation (\ref{F3})).

  Next, we derive slow-roll parameters, $\e  \equiv -{\dot{H}}/{H^2}\,,
 \e_V \equiv  \l({M^2_\p}/{2}\r) \l({V_{\phi}}/{V}\r)^2
  \,,\delta    \equiv  -{\ddot{\phi}}/{(H \dot{\phi})}\,,
\eta  \equiv {\dot{\e}}/{(H \e)}$  and   $\eta_{V}    \equiv  M^2_\p | {V_{\phi \phi}}/{V}|$ of this model. These recover slow-roll parameters in standard GR case when $\ve = 1, c = 0$ and $\k = 0$.
The spectral index $n_{\text{s}}$, the tensor-to-scalar ratio $r$ and expression of e-folding number $N$  are derived. Constraint on single field inflation
obtained from Planck 2018 CMB data \cite{Planck:2018vyg} gives $ n_{\text{s}} = 0.965 \pm  0.004$  and with
 BICEP-Keck 2015 data on B-mode polarization, $r_{0.002} < 0.06$ at 95 \% CL.
We are interested in only non-phantom case with  $V(\phi) > 0$.  Scalar potentials  considered are in form of
$V = V_0 \phi^n $ with $n = 2, 4$ and  $V = V_0 \exp{(-\beta \phi)}$. Amount of e-folding number is taken to be $N \geq 60$  so that the horizon problem is  evaded. For $n = 2$,
in case of $\alpha(\k, c) \equiv  \ve - {\k V}/[M^2_\p (1-c^2)]  >  0 $, the allowed range of $\k$ is approximately,
$
-0.206  \,<\, \k V_0 \, \leq  0.002 \,.
$
and in case of $\alpha < 0$,  the allowed range is
$ 0 < \k V_0\,$.
Combined parametric plots of $\k$ and $\phi$  for these potentials with constraints from $n_{\text{s}}$  and $r$ upper bound obtained from CMB data \cite{Planck:2018vyg} reveal that they are not overlapping.
Only $N \geq 60 $  region can overlap with either $n_{\text{s}}$   allowed region or $r$ allowed region separately.
Holographic contribution has significant effects in the parametric plots. The plots do not allow any value of $\k$  to be valid for $n_{\text{s}}$ constraint and  $r$  constraint at the same time. Therefore, for $c = 0$, our result confirms the shortcomings of the non-holographic NMDC inflation case reported earlier in
\cite{Tsujikawa:2012mk, Yang:2015pga, Avdeev:2022ilo}. Moreover, for $0 < c < 1$, the holographic NMDC inflation is hence not compatible with inflation for all potentials considered. We conclude that the $0 \leq c < 1$ cases are ruled out for NMDC inflation.

The model may be considered in the late universe as sources of dark energy. Phase portraits in figure \ref{fig2} show that NMDC contribution gives complicated phase portraits as well as alters their acceleration regions.
Holographic part can affect the acceleration region only in presence of the  NMDC field.  We see qualitatively that holographic part does not affect anatomy of the dynamics as it only changes shape of the acceleration regions and pattern of the phase trajectories.
On the other hand, the NMDC part changes anatomy of the dynamics, i.e. the NMDC gives new late-time attractor trajectories locating in acceleration regions.

At late time, variation of gravitational constant can be used to constrain our model.
Constraint from gravitational-wave standard sirens and supernovae data, $\dot{G}/G|_{t_0} \lesssim 3\times10^{-12} \,  \text{year}^{-1}$ \cite{Zhao:2018gwk} is considered in our work. Variation of gravitational constant in this model is given by,
$
{\dot{\tilde{G}}_{\text{eff}}}/{\tilde{G}_{\text{eff}} }   \,=\,   {\dot{G}_{\text{eff}}}/{G_{\text{eff}} }
\,   \simeq  \,      {-3 \kappa \dot{\phi}\ddot{\phi}}/{M^2_{\p}}\,. $  The holographic part contributes implicitly via $\dot{\phi}$ and $\ddot{\phi}$. The constraint relation is,
$
10^{-12} \,  \text{year}^{-1}   \,  \gtrsim  \,      {- \kappa} \dot{\phi}\ddot{\phi}/{M^2_{\p}}\,.
$
Using slow-roll approximation, for each type of potential, we find late-time scalar field solutions which are used to derive constraint conditions of the NMDC coupling, $\k$ of which the value is presented  in table \ref{TableG}. We find that when the holographic effect, $c^2$  is greater, the lower bound of $\k$ is lifted up for all types of potential.
For instance, $ 0.0072/m^2  \lesssim \k $ for $c=0$ and  $ 0.0096/m^2  \lesssim \k $ for $c=0.5$ for $V=(1/2)m^2 \phi^2$.

In summary,
the non-holographic NMDC inflation and the holographic NMDC inflation with Bekenstein-Hawking entropy and Hubble scale cutoff with $0 \leq c < 1$  are
ruled out for power-law and decreasing exponential potentials. We suspect that using apparent horizon cutoff, i.e. the case $k \neq 0$ would not make significant difference in the power spectrum index and tensor-to-scalar ratio. NMDC effect gives new late-time attractor trajectories in scalar field phase portraits.   At late time, taking role of dark energy, the coupling $\k$  is viable within positive range with lower bounds reported here.  Kinematic behaviors such as late expansion of this model and the analysis when $c = 1$ are interesting for future works.

%%%%%%%%%%%%%%%%%%%%%%%%%%%%%%%%%%%%%%%

%%%%%%%%%%%%%%%%%%%%%%%%%%%%%%%%%%%%
%%%%%%%%%%%%%%%%%%%%%%%%%%%%%%%%%%%%%%%%%%%%%%%%%%%%%

\section*{Acknowledgments}
 This research project is supported by Mahidol University (grant no. MRC-MGR 04/2565).
 PB  is supported by a DPST scholarship.  CK is funded by University of Phayao research grant no. FF64-RIM07.

%%%%%%%%%%%%%%%%%%%%%%%%%%%%%%%%%%%%%%%%%%%%%%%

\end{document}